\documentclass[sigconf,screen, authorversion]{acmart}

\AtBeginDocument{%
  \providecommand\BibTeX{{%
    \normalfont B\kern-0.5em{\scshape i\kern-0.25em b}\kern-0.8em\TeX}}}

\acmConference[ARES 2022]{International Conference on Availability, Reliability and Security}{August 23--26, 2022}{Vienna, Austria}

\usepackage{graphicx}
\usepackage[misc,geometry]{ifsym}
\usepackage{booktabs}
\usepackage{multirow}
\usepackage{pgfplots}
\pgfplotsset{compat=1.16}

\usepackage{listings}
\usepackage{pgf-pie}
\usepackage{booktabs}
\usepackage{dot2texi}
\usepackage{diagbox}
\usepackage{makecell}

\usepackage[T1]{fontenc}
\usepackage{url}
\usepackage[keeplastbox]{flushend}
\usepackage{multicol}
\usepackage{colortbl}
\usepackage{array}
\usepackage{amsmath}
\usepackage{svg}

\usepackage{balance}

\usetikzlibrary{shapes, arrows, patterns, automata, positioning}

\copyrightyear{2022}
\acmYear{2022}
\setcopyright{acmlicensed}\acmConference[ARES 2022]{The 17th International Conference on Availability, Reliability and Security}{August 23--26, 2022}{Vienna, Austria}
\acmBooktitle{The 17th International Conference on Availability, Reliability and Security (ARES 2022), August 23--26, 2022, Vienna, Austria}
\acmPrice{15.00}
\acmDOI{10.1145/3538969.3544458}
\acmISBN{978-1-4503-9670-7/22/08}

\begin{document}

\title[Current Challenges of Cyber Threat and Vulnerability Identification Using Public Enumerations]{Current Challenges of Cyber Threat and Vulnerability Identification Using Public Enumerations}

\author{Lukáš Sadlek}
\affiliation{%
    \institution{Masaryk University}
    \city{Brno}
    \country{Czech Republic}
}
\email{sadlek@mail.muni.cz}
\orcid{0000-0003-2577-6633}

\author{Pavel \v{C}eleda}
\affiliation{%
    \institution{Masaryk University}
    \city{Brno}
    \country{Czech Republic}}
\email{celeda@ics.muni.cz}
\orcid{0000-0002-3338-2856}

\author{Daniel Tovar\v{n}\'ak}
\affiliation{%
  \institution{Masaryk University}
  \city{Brno}
  \country{Czech Republic}
}
\email{tovarnak@ics.muni.cz}
\orcid{0000-0002-7206-5167}

\renewcommand{\shortauthors}{L. Sadlek et al.}

\begin{abstract}
    Identification of cyber threats is one of the essential tasks for security teams. Currently, cyber threats can be identified using knowledge organized into various formats, enumerations, and knowledge bases.
    This paper studies the current challenges of identifying vulnerabilities and threats in cyberspace using enumerations and data about assets. 
    Although enumerations are used in practice, we point out several issues that still decrease the quality of vulnerability and threat identification.
    Since vulnerability identification methods are based on network monitoring and agents, the issues are related to the asset discovery, the precision of vulnerability discovery, and the amount of data. On the other hand, threat identification utilizes graph-based, nature-language, machine-learning, and ontological approaches. 
    The current trend is to propose methods that utilize tactics, techniques, and procedures instead of low-level indicators of compromise to make cyber threat identification more mature. 
    Cooperation between standards from threat, vulnerability, and asset management is also an unresolved issue confirmed by analyzing relationships between public enumerations and knowledge bases.
    Last, we studied the usability of techniques from the MITRE ATT\&CK knowledge base for threat modeling using network monitoring to capture data. Although network traffic is not the most used data source, it allows the modeling of almost all tactics from the MITRE ATT\&CK.
\end{abstract}

\ccsdesc[500]{Security and privacy~Network security}
\ccsdesc[500]{Security and privacy~Vulnerability management}
\ccsdesc[300]{Networks~Network monitoring}

\keywords{cyber threat intelligence, threat identification, vulnerability identification, CVE, CAPEC, MITRE ATT\&CK}

\maketitle

\section{Introduction}
Threat identification is the task of revealing any possible event jeopardizing organization’s assets. Several enumerations and knowledge bases support such a use case by defining vocabulary for specific areas of cybersecurity~\cite{pawlinski2014standards}. \textit{Common Vulnerabilities and Exposures} (CVE)~\cite{cve} define how vulnerability identifiers are assigned to discovered vulnerabilities. \textit{Common Weakness Enumeration} (CWE)~\cite{cwe} categorizes common weaknesses, while \textit{Common Platform Enumeration} (CPE)~\cite{cpe} defines a way of assigning standardized identifiers to classes of technological assets. Last, \textit{Common Attack Pattern Enumeration and Classification} (CAPEC)~\cite{capec} classifies common attack patterns. A related knowledge base \textit{MITRE ATT\&CK}~\cite{attack} provides a matrix of adversarial tactics and techniques.

These enumerations are used within the \textit{cyber threat intelligence} (CTI) that provides knowledge about goals, methods, previous attacks, and current attack possibilities of adversaries in cyberspace.
Even though more than half of the \textit{SANS CTI survey} respondents use CTI for threat and vulnerability management~\cite{brown2019evolution}, only 39\% of organizations knew that they were vulnerable prior to the data breach based on Ponemon Institute's study~\cite{ponemon2019}. These statistics show that current approaches for threat and vulnerability identification may not be perfect.

These enumerations belong to the threat, vulnerability, and asset management. Therefore, the interoperability of these managements should support the identification of cyber threats. For example, \textit{NIST Cybersecurity Framework} recommends several activities supporting risk evaluation of cyber threats caused by vulnerable assets, optionally using shared CTI~\cite{nistcyberframework}, such as public enumerations. Data from public enumerations need to be enriched with data about the organization's assets obtained from network monitoring, network scanning, and application and system logs.

In this paper, we focus on identifying threats and vulnerabilities in cyberspace using the aforementioned public enumerations and data about assets. For the simplicity of expression, we also consider the ATT\&CK knowledge base when using term enumerations in this paper. We propose the following research questions:

\begin{enumerate}
    \it
    \item What are the current challenges of vulnerability and cyber threat identification using enumerations and data about assets?
    \item What is the usability of MITRE ATT\&CK for threat modeling when only network monitoring is used as a source of data?
    \item What is the interoperability of public enumerations using references between their entries?
\end{enumerate}

In the first research question, we summarize the current state of the art. We focus on methods that identify enumeration entries, excluding methods that discover only some entries using their related detection methods. 
In the second research question, we study whether the well-known MITRE ATT\&CK knowledge base can be used for cyber threat identification using network monitoring. Finally, the third research question evaluates the interoperability of the enumerations using their references.

The paper is divided into six sections. First, we describe related work in Section \ref{sub:rel_work}. Section \ref{sec:vul_id} contains state of the art for vulnerability identification and challenges of the current state. Section \ref{sec:threat_id} describes the identification of threats using public enumerations and knowledge bases, including the current challenges in this area. The following Section \ref{sec:analysis} describes the results of the analysis of public enumerations. Last, Section \ref{sec:conclusion} concludes the paper.

\section{Related Work} \label{sub:rel_work}
The related work consists of several parts related to the threat and vulnerability identification. The first part covers enumerations actively used in this research area. The following part introduces the position of these enumerations within the field of the \textit{cyber threat intelligence} (CTI). The last part justifies that related work did not cover this area of research.

\subsection{Security Enumerations}
\begin{figure}[b]
\begin{center}
\includegraphics[width=8.5cm]{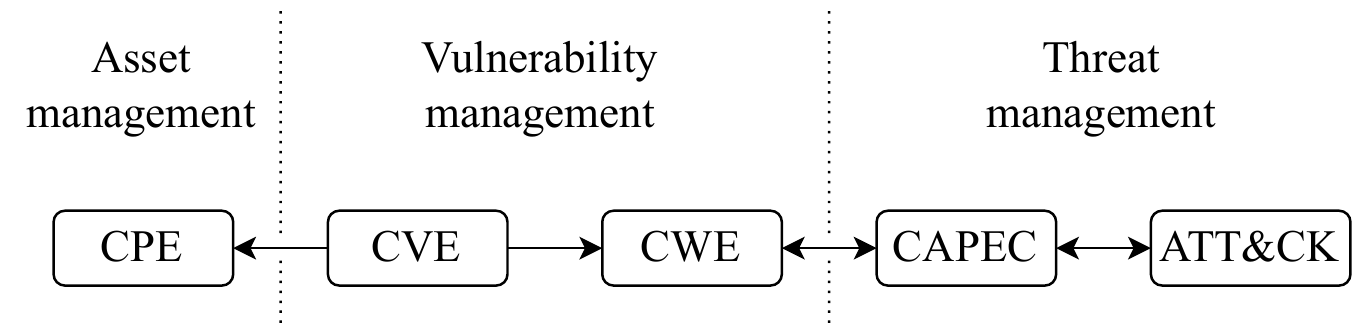}
\end{center}
\caption{Mapping of data entities about assets, vulnerabilities, and threats.} \label{fig:refer_new}
\end{figure}

The enumerations considered in this paper belong to different parts of the security operations management -- asset, vulnerability, and threat management (see Figure \ref{fig:refer_new}).
An enumeration belonging to asset management is called the \textit{Common Platform Enumeration} (CPE)~\cite{cpe}. 
The well-known enumerations that belong to vulnerability management are the \textit{Common Vulnerabilities and Exposures} (CVE), listing discovered vulnerabilities, and the \textit{Common Weakness Enumeration} (CWE), listing commonly appearing weaknesses. 
\textit{Common Attack Pattern Enumeration and Classification} (CAPEC)~\cite{capec} and \textit{MITRE ATT\&CK}~\cite{attack} belong to threat management.

The CPE is a standard for naming classes of products -- hardware devices, operating systems, and applications. These classes with other product information are specified within a formatted string called the CPE match string used in data feeds from the \textit{National Vulnerability Database (NVD)}~\cite{nvd}.

\begin{figure}[h]
\fbox{\begin{minipage}{0.9\linewidth}
\small
\textbf{cpe:2.3}:<part>:<vendor>:<product>:<version>:<update>:<edition>:
:<language>:<sw\_edition>:<target\_sw>:<target\_hw>:<other>
\end{minipage}}

\fbox{\begin{minipage}{0.9\linewidth}
\small
\textbf{cpe:2.3}:o:debian:debian\_linux:11.0:*:*:*:*:*:*:*\\
\textbf{cpe:2.3}:a:apache:log4j:2.0:rc1:*:*:*:*:*:*
\end{minipage}}
\caption{CPE match string and its example content.} \label{cpematchstring}
\end{figure}

The current version, CPE 2.3, uses syntax expressed in Figure~\ref{cpematchstring}~\cite{cpe}. The most important parts are at the beginning of the string. The \textit{part} section has three possible values: \textit{a} for applications, \textit{o} for operating systems, and \textit{h} for hardware. The \textit{vendor} expresses the name of the product's vendor. The product name is filled in the \textit{product} section and followed by the product's version. The CPE match string may also contain the product's update, edition, and other parts~\cite{cpe}. The syntax allows specifying special characters, e.g., an asterisk is used to denote \textit{ANY} value in the appropriate part of the formatted string.

The \textit{Common Vulnerabilities and Exposures} (CVE)~\cite{cve} is a list of vulnerabilities published in the \textit{National Vulnerability Database} (NVD)~\cite{nvd} maintained by MITRE. Each CVE entry contains an identification number, description, and URL references to sources where the vulnerability was published, e.g., on vendor websites. The NVD stores CVE vulnerabilities also with their weaknesses (CWE ID)
and vulnerable products (CPE). 

An example of a CPE match string for CVE vulnerability is the \textit{log4j's} match string associated with the \textit{Log4Shell} vulnerability (see Figure \ref{cpematchstring}). The NVD~\cite{nvd} also specifies CPE configurations consisting of several CPEs grouped by keywords \textit{AND} and \textit{OR}. In such a way, the NVD specifies that CVE is present, e.g., on a specific operating system \textit{AND} application installed on it. Using the \textit{OR} keyword specifies several vulnerable CPE configurations.

The \textit{Common Weakness Enumeration} (CWE)~\cite{cwe} lists common weaknesses in software or hardware products. This enumeration is organized as a tree with well-known examples of weaknesses, such as \textit{CWE-81 (XML Injection)} and \textit{CWE-94 (Code Injection)}. The entries also contain additional information, such as descriptions and mitigation methods. Related CWEs are assigned to each vulnerability in the NVD, e.g., \textit{Log4Shell} vulnerability (CVE-2021-44228) is categorized as \textit{CWE-502} (Deserialization of Untrusted Data). CWE also contains a mapping from \textit{OWASP Top Ten 2021}.

CAPEC describes attack patterns and ATT\&CK adversarial tactics and techniques. CAPEC is organized as a hierarchical tree of categories, e.g., \textit{CAPEC-125 (Flooding)} is a parent of \textit{CAPEC-482 (TCP Flood)}, \textit{CAPEC-486 (UDP Flood)}, and other subcategories of the flood. On the contrary, MITRE ATT\&CK is organized as a matrix where columns are represented by tactics and rows by techniques. Examples of ATT\&CK techniques are \textit{Phishing (T1566)} and \textit{Brute Force (T1110)}. Examples of references between data entries from Figure \ref{fig:refer_new} are \textit{CAPEC-98} and the ATT\&CK technique with ID \textit{T1566}. Both of them reference each other and express phishing.

The \textit{Structured Threat Information Expression} (STIX) is a format for expressing and serialization of CTI~\cite{stixpart1}. STIX represents a threat as a graph with CTI entities (e.g., attack pattern, vulnerability, malware, and course of action) and their relationships. The STIX allows storing entries from the enumerations in its objects. For example, \textit{attack pattern} objects may contain CAPEC and ATT\&CK entries, while \textit{vulnerability} objects often contain CVE vulnerabilities.

\textit{Sauerwein et al.}~\cite{Sauerwein2019} conducted a triangulation study about public security data sources. According to the authors, information about vulnerabilities and attacks or vulnerabilities and threats often appears in one data source. They found out that CVE and STIX are two primary standards that are used. CVE identifiers appear almost in every second data source, CWE in 13\% of them,
while the other standardized enumerators rarely appear~\cite{Sauerwein2019}. Therefore, the task of cyber threat and vulnerability assessment using the enumerations is not trivial and has its rationale. A summary of mentioned enumerations is in Table \ref{tab:stand_table}.

\begin{table}[t]
    \centering
    \caption{Comparison of structure and sources of mentioned enumerations.}
    \label{tab:stand_table}
    \begin{tabular}{ m{1.3cm} | m{2.2cm} | m{1.4cm} | m{2.0cm} }
    \toprule
    \multirow{2}{1.3cm}{\textbf{Source}} & \multirow{2}{2.2cm}{\textbf{Structure}} & \textbf{Free-Text} & \multirow{2}{2.0cm}{\textbf{Data Source}} \\
     & & \textbf{Fields} & \\ \midrule
    \multirow{2}{1.3cm}{CVE} & flat & \multirow{2}{1.4cm}{partially} & \multirow{2}{2.0cm}{NVD}\\
     & (list) & & \\ \hline
    \multirow{2}{1.3cm}{CPE} & hierarchical & \multirow{2}{1.4cm}{no} & \multirow{2}{2.0cm}{NVD}\\
     & (string) & & \\ \hline
    \multirow{2}{1.3cm}{CWE} & hierarchical & \multirow{2}{1.4cm}{partially} & \multirow{2}{2.0cm}{MITRE website}\\
     & (tree) & & \\ \hline
    \multirow{2}{1.3cm}{CAPEC} & hierarchical & \multirow{2}{1.4cm}{partially} & MITRE website,\\ 
     & (tree) & & Github \\ \hline
    \multirow{2}{1.3cm}{ATT\&CK} & hierarchical & \multirow{2}{1.4cm}{partially} & \multirow{2}{2.0cm}{Github}\\
     & (matrix) & & \\
	\bottomrule
    \end{tabular}
\end{table}

\subsection{Position of Enumerations in Cyber Threat Intelligence} \label{sec:position}
Enumeration entries can be categorized into several CTI categories according to their maturity. CTI is generally related to indicators of compromise (IOCs) that can be divided into atomic, computed, and behavioral~\cite{hutchins2011}. Atomic indicators cannot be divided anymore, e.g., IP addresses or CVE IDs. On the contrary, computed indicators are literally computed from the captured data, e.g., file hash values. Lastly, behavioral indicators describe the attacker's behavior using atomic and computed indicators~\cite{hutchins2011}. E.g., ATT\&CK and CAPEC IDs describing the attacker's activity are behavioral indicators.

Relationships among distinct parts of CTI can also be expressed using the \textit{Pyramid of Pain (PoP)}~\cite{bianco2013} and the \textit{Detection Maturity Level (DML)} model~\cite{stillions2014} or its extended version~\cite{Mavroeidis2017}. The particular PoP levels express the attacker's difficulties when the defense operates with more sophisticated indicator types. Similarly, the DML model expresses the organization's maturity in applying different indicator types, from the most low-level to the most abstract. The DML model and the PoP are similar, but the PoP divides atomic indicators into three different levels (see Table \ref{tab:comparison}). Tactics, techniques, and procedures (TTPs) that can be obtained, e.g., from MITRE ATT\&CK, express the most mature indicators used for the security defense.

\begin{table}[t]
    \centering
    \caption{Comparison of levels from DML model and PoP.}
    \label{tab:comparison}
    \begin{tabular}{>{\centering\arraybackslash}m{0.3cm}|m{3.7cm}|m{3.2cm}}
    \toprule
    & \textbf{Detection Maturity Level} & \textbf{Pyramid of Pain} \\ \midrule
        \multirow{3}*{1} & \multirow{3}*{atomic indicators} & domain names \\
         & & IP addresses \\
         & & hash values\\ \hline
         2 & host \& network artifacts & network \& host artifacts \\ \hline
        3 & tools & tools \\ \hline
        4 & procedures & \multirow{3}*{TTPs} \\
        5 & techniques &  \\
        6 & tactics &  \\ \hline
        7 & strategy & \multirow{3}*{------} \\ 
        8 & goals & \\
        9 & identity &  \\
	\bottomrule
    \end{tabular}
\end{table}

\subsection{Cyber Threat Intelligence Surveys} \label{subsec:ex_surveys}
In this section, we justify that our research area is not described sufficiently by the related work since our paper's limitation is an absence of methodological literature search.
In general, the existing surveys study the area of CTI without focusing on identifying cyber threats and vulnerabilities using the enumerations. 

\textit{Tounsi and Rais}~\cite{Tounsi2018} surveyed issues and emerging trends of technical threat intelligence. They pointed out that the traditional approaches in this area are not sufficient to cope with advanced persistent threats, zero-day threats, composite threats, or polymorphic threats. \textit{Mavroeidis and Bromander}~\cite{Mavroeidis2017} evaluated CTI taxonomies, sharing standards, and ontologies. They introduced their CTI model since there was no complete ontology covering all types of CTI data. \textit{Wagner et al.}~\cite{Wagner2019} surveyed the current state of CTI sharing. They claim that it is challenging to participate in CTI sharing because of insufficient models and tools. 

Threat modeling develops threat representation and analyzes adversarial possibilities. \textit{Bodeau et al.}~\cite{Bodeau2018} prepared a technical report about cyber threat modeling comparing characteristics of threat models and frameworks. \textit{Xiong and Lagerström}~\cite{Xiong2019} conducted a systematic literature review about threat modeling. They found that threat modeling is diverse, but most activities are manually accomplished. 
A promising research direction is to connect threat models with threat and vulnerability databases.

Other surveys may seem to be related to this paper but do not focus on identifying vulnerabilities and cyber threats to the organization's assets. \textit{Le et al.}~\cite{le2021survey} surveyed data-driven vulnerability assessment using machine learning and deep learning. They focused on the prioritization and the prediction of exploitation, impact, severity, type, and other properties of vulnerabilities. \textit{Rahman et al.}~\cite{rahman2021attackers} provided a survey of threat intelligence extraction from text. The primary goal was to extract important entities, not to identify threats in the organization. Therefore, only some papers listed in the survey are relevant to our topic. \textit{Kaloudi and Li}~\cite{Kaloudi2020ai} surveyed the cyber threat landscape related to artificial intelligence but without focusing on the enumerations.

\section{Vulnerability Identification} \label{sec:vul_id}
Vulnerability identification using public enumerations is a task to reveal the presence of a published vulnerability in the organization. The task can be accomplished in two manners. The first one is based on asset discovery and vulnerability databases. The second one uses published exploits to reveal vulnerable systems. It is more reliable and is often used as \textit{vulnerability validation} after applying the first one.

\subsection{State of the Art}
Vulnerability identification methods based on asset discovery may obtain information about assets from passive network monitoring, active network monitoring, and agents. Passive network monitoring consists of capturing network traffic at some observation point~\cite{rick2014flow} without any modification of traffic. On the contrary, active network monitoring targets artificial requests to network services using network scanners~\cite{BouHarb2014}. Agent-based approaches use software checking installed applications on devices and their versions. In the following text, we will focus on vulnerability identification approaches built upon these data collection methods.

The task of identifying CVE vulnerability using data about assets can be divided into two subtasks -- construct the CPE match string and find the appropriate CVE from the NVD using the constructed string.
The first subtask can be accomplished using banner grabbing and OS and network service fingerprinting if the network monitoring is used.

\textit{Banner grabbing} is a technique of obtaining responses from services running on opened ports when a network connection is initiated. This response is parsed to extract parts of the CPE match string in Figure \ref{cpematchstring}. Usually, such a method can provide part, vendor, product, version, and sometimes even other fields of the string.

\textit{O'Hare et al.}~\cite{Hare2019} used metadata obtained from the Censys scanner that parses service banners to obtain the list of CPEs representing vulnerable services. \textit{Genge and En\u{a}chescu}~\cite{Genge2016} used the results of Shodan queries. They extracted vendor names, product names, and product versions from service banners to identify relevant vulnerabilities from the NVD. Shodan and vulnerabilities from the NVD were also used by \textit{Samtani et al.}~\cite{Samtani2016}.

\textit{OS and network service fingerprinting} are other options for constructing the CPE match string. OS fingerprinting captures network connection properties, such as TCP window size and Time to Live (TTL), to infer the device's operating system. Passive OS fingerprinting from IP flow was used by \textit{Laštovička et al.}~\cite{Lastovicka2020} to enumerate vulnerabilities in large networks. 

Banner grabbing and fingerprinting approaches are often implemented in network scanners, e.g., Nmap~\cite{nmap} and Shodan. The Nmap scanner outputs the CPE string containing vendor name, product name, and product version of scanned network services.

The network monitoring can be substituted by system and application logs to identify products installed on devices. An approach similar to agent-based approaches was proposed by \textit{Gawron et al.} They utilized Unix system logs, Windows log events, web server logs, and proxy logs to extract CPE identifiers of applications~\cite{Gawron2017}.

The second subtask is to determine CVE using the constructed string. Related work mentions several options for this subtask. Linear search through the CPE dictionary is the most straightforward option. For example, sequential lookup and Levenshtein distance were used by \textit{O'Hare et al.}~\cite{Hare2019}. \textit{Genge and En\u{a}chescu}~\cite{Genge2016} introduced a more mature approach. They used the hash tables for storing possible CPE names and a tree for hierarchically storing version numbers. Bipartite graphs were then used to model the relationship between CVEs and related CPEs, while hyper edges modeled CPE configurations (see Section \ref{sub:rel_work}).

An approach that does not use data about assets but checks a published exploit or presence of weakness is implemented in vulnerability scanners. Some scanners (e.g., Arachni) can identify weaknesses such as XSS, SQL, and code injections, and output discovered weaknesses as CWE and CAPEC entries. Nessus identifies network services, operating systems, and vulnerabilities~\cite{nessus}. Metasploit is an example tool~\cite{metasploit} that validates the presence of vulnerabilities using exploits from Exploit DB~\cite{exploitdb}. Some of these exploits even have their CVE ID assigned.

\subsection{Challenges}
Challenges of vulnerability identification using public enumerations arise from the relationship of vulnerability management with asset management, shortcomings of current methods, and a large amount of data.

\subsubsection*{\textbf{Asset Management:}}
Vulnerability identification is associated with situational awareness about vulnerable assets in the organization. However, according to Edgescan's 2020 Vulnerability statistics report, most professionals do not know about all organization's assets, e.g., web applications or endpoints~\cite{Edgescan2020}. When methods based on asset discovery are used, it may cause some organizations to successfully patch vulnerabilities on known assets while leaving the unknown completely unpatched~\cite{dbir2020}.

\subsubsection*{\textbf{Vulnerability Discovery Precision:}}
Another issue of vulnerability discovery is a lot of false positives. Methods that have extensive coverage (e.g., based on OS fingerprinting) can be imprecise or will not reveal sufficient details~\cite{Lastovicka2020}. It causes employees to spend much time validating vulnerabilities~\cite{Edgescan2020}, e.g., using vulnerability scanners that can use automated vulnerability exploits to confirm true positives.

In our scope, this issue is caused by constructing a CPE string that contains many \textit{ANY} values. \textit{Gawron et al.}~\cite{Gawron2017} can often extract only an application name and a version from logs for the CPE match strings (see Figure \ref{cpematchstring}). On the contrary, \textit{Laštovička et al.}~\cite{Lastovicka2020} used active scanning and passive monitoring that can determine only the first parts of the match string up to the version part.

Evaluation can be found in several papers. Examples of discussed approaches are implemented in Scout, which has 73 \% precision \cite{Hare2019}, and ShoVAT, which outputs 7.77 \% false positives \cite{Genge2016}. Researchers usually use the results of other vulnerability scanners or approaches as ground truth for their methods \cite{Lastovicka2020, Hare2019, Genge2016}.

\subsubsection*{\textbf{Amount of Data:}}
The NVD currently contains more than 170 thousand vulnerabilities. More than 8000 entries were added to the CPE dictionary during each month of 2021~\cite{cpe_statistics}. Currently, implemented approaches are still efficient enough to be used in practice. However, timely aspects and asset landscape visibility are becoming more critical with the increasing amount of published vulnerabilities.

\subsubsection*{\textbf{Implementation of CPE Specifications:}}
CPE has several specifications that standardize CPE naming, CPE dictionary, CPE applicability language, and CPE name matching~\cite{tovarnak2021graph}. However, only specification for CPE naming of products~\cite{cpe} is usually used in research papers. These research papers implement custom algorithms for CPE matching. A graph-based approach that conforms to all four specifications was provided in~\cite{tovarnak2021graph}.

\subsection{Possible Research Directions}
We identified two possible research directions based on the current state of the art. Since commercial solutions for vulnerability management consider vulnerability identification one of its essential tasks, their creators are motivated to advance the current state. However, we still need approaches backed by scientific evaluations.

\subsubsection*{\textbf{Interoperability:}}
The first possible research direction is to create a functioning system with good network visibility by unifying partial approaches. This research direction has its rationale since vulnerability identification approaches have some advantages. They were evaluated by \textit{Laštovička et al.}~\cite{Lastovicka2020} who showed that the passive network monitoring reveals more vulnerable hosts assuming a particular OS fingerprinting method can process encrypted traffic. However, active network monitoring has better accuracy because of sending specially crafted probes to identify selected services. Commercial solutions also unify the mentioned approaches (active monitoring, passive monitoring, and agents) to achieve higher accuracy of vulnerability identification for customers~\cite{QualysVM, TenableVM}.

\subsubsection*{\textbf{Current IT Environments:}}
Since current IT environments consist of on-premises and cloud assets, networks lose their traditional perimeters. Listed academic literature also does not focus on other novel aspects of IT environments, e.g., encrypted network traffic that hinders the use of some methods. However, commercial solutions for vulnerability management can cope with these environments and work with on-premises assets, cloud assets, and mobile endpoints~\cite{QualysVM, TenableVM}. Their disadvantage is the missing detailed insight into these paid tools.

\section{Threat Identification} \label{sec:threat_id}
Cyber threat identification belongs to essential tasks of threat management. For example, Cisco assigns capabilities such as threat intelligence services
to the fourth generation of security operations centers~\cite{Muniz2016}. Cyber threat identification using the enumerations determines whether some categorized threat could materialize according to the current security posture. Naive identification of such threats correlates indicators of compromise with data obtained from the organization. However, the use of enumerations usually leads to more sophisticated methods that utilize \textit{Tactics, Techniques, and Procedures (TTPs)}. 

\subsection{State of the Art}
Enumeration entries can appear in CTI reports~\cite{Husari2017}, within data models of cyber threats~\cite{Mavroeidis2018, Zhao2020}, and be extracted from system and application logs~\cite{Milajerdi2019Holmes}. Peer-reviewed literature provides several types of approaches that use the enumerations. Graph-based approaches can model entities (e.g., events) and their relationships~\cite{Milajerdi2019Holmes}. Natural Language Processing (NLP) processes human-readable threat intelligence~\cite{Husari2017, Zhao2020} and can accomplish additional tasks with machine learning (ML), e.g., classification~\cite{Noor2019}. Last, ontological approaches help model the CTI domain~\cite{Mavroeidis2018, Qamar2017}.

Graph-based approaches are helpful for modeling sequences of steps related to TTPs. These steps may be accomplished simultaneously with other legitimate activities of users, and attackers may skip or add some steps. Therefore, the rational method is to search for similarities with malicious TTPs. Such approaches may provide more granular results compared to simple detection mechanisms.

A graph-based approach for the detection of Advanced Persistent Threats (APTs) was proposed by \textit{Milajerdi et al.}~\cite{Milajerdi2019Holmes} based on a provenance graph from system audit logs and a high-level scenario graph containing APT's activities (TTPs) mapped to the cyber kill chain. The TTPs are found in the provenance graph and mapped to their kill chain phases. Each of the TTPs has its assigned severity from CAPEC. The final score is computed based on all kill chain phases to distinguish attacks with sufficient confidence.

The NLP is applied to automatically extract the data entities (e.g., TTPs, IOCs, domain tags) from human-readable CTI texts. However, this approach also requires a custom or a standardized taxonomy of entities that can be extracted, e.g., CAPEC and ATT\&CK for TTPs~\cite{Husari2017}. This area contains a lot of scientific literature that describes extraction methods~\cite{rahman2021attackers} but does not provide means to identify relevant threats to the protected assets. \textit{Zhao et al.}~\cite{Zhao2020} extracted CTI from social data using NLP and a convolutional neural network to identify CVE and other indicators from the CTI. Relevant threats are determined according to domain tags that describe where the organization belongs, e.g., government.

Machine-learning classification can be applied when extracted CTI entities can be ambiguous. \textit{Noor et al.}~\cite{Noor2019} trained naive Bayes classifier processing incident descriptions to determine TTPs from ATT\&CK. A semantic network from threats and the TTPs allows determining relevant threats according to the host and network artifacts belonging to the second level of DML in Section \ref{sec:position}.

Ontological approaches are usually based on a representation of CTI using standards, taxonomies, threat reports, and network architecture. Populated ontologies allow inferring new relationships among their entities. The CTI ontology was used by \textit{Mavroeidis and Jøsang}~\cite{Mavroeidis2018} for threat hunting from Sysmon logs. The ontology contains, e.g., threat actors, attack patterns, and indicators of compromise. It can reveal the attacker's goals and suitable countermeasures. \textit{Qamar et al.}~\cite{Qamar2017} created ontology that maps threats to the network based on STIX reports, CVE, and network architecture. The inference rules determine relevant threats and identify affected assets based on the CTI, its TTPs, indicators, and vulnerabilities.

\subsection{Challenges}
Challenges of threat identification using CTI arise from the relationship of threat management with asset and vulnerability managements, a large amount of data, unstructured data, and shortcomings of current methods.

\subsubsection*{\textbf{Unstructured CTI Reports:}}
Unstructured text of CTI reports causes difficulties with automated processing. One of the issues is to extract relevant entities~\cite{rahman2021attackers}, e.g., an attacker's name, techniques, and exploited vulnerabilities. Standardized enumerations and knowledge bases provide taxonomies of obtainable output, e.g., vulnerability ID and name of an attack technique. Therefore, researchers often utilize text mining and NLP~\cite{Zhao2020}.

\subsubsection*{\textbf{Lack of Visibility and Amount of Data:}}
Organizations usually obtain data from several external sources to identify cyber threats~\cite{brown2019evolution}. Therefore, the organizations cope with a large amount of CTI obtained from different sources and with different formats. It is also necessary to add internal context about assets. As a result, researchers aim to provide means of determining relevant threats~\cite{Zhao2020, Noor2019, Qamar2017}.

\subsubsection*{\textbf{Maturity of Methods:}}
Naive methods correlate suspicious IOCs from CTI reports with data from the organization. However, such a method’s usability is limited because the attacker can easily and quickly change these indicators, e.g., the validity of IP addresses quickly decreases after one day~\cite{Tounsi2018}. Therefore, researchers introduce methods that support TTPs~\cite{Mavroeidis2018, Milajerdi2019Holmes, Zhao2020, Noor2019, Qamar2017} to elevate the methods to higher levels of DML. Researchers usually evaluate methods on past CTI reports~\cite{Qamar2017} or prepare some dataset containing threat-related activity, e.g., from a cybersecurity exercise~\cite{Milajerdi2019Holmes}.

\subsection{Possible Research Directions}
We identified two possible research directions related to this topic. The first one is the same as for the vulnerability identification, and the second one is related to machine learning. Both may be partially related to the current topic of SOAR (\textit{Security Orchestration, Automation, and Response}) platforms~\cite{gartnersoar}.

\subsubsection*{\textbf{Interoperability of Partial Solutions:}}
The task is to research how to combine partial solutions to provide a functional system able to identify a wide range of cyber threats using the enumerations. It would be necessary to use several data sources~\cite{gartnersoar} and methods that accomplish more than a direct combination of security alerts.

\subsubsection*{\textbf{Machine Learning for Threat Identification:}}
Machine learning provides several methods that can be applied for threat identification using the enumerations, such as classification and clustering. Some commercial tools apply machine learning within threat management, e.g., graph-based machine learning~\cite{alienvaultusm} and artificial intelligence~\cite{IBMQRadar}. However, the use of artificial intelligence is a threat itself~\cite{Kaloudi2020ai}.

\section{Analysis of Enumerations} \label{sec:analysis}
This section provides results that answer the second and the third research question. First, we parsed the ATT\&CK matrix expressed in the STIX format from the MITRE's CTI repository~\cite{mitrecti}. This matrix contained tactics, techniques, data sources, countermeasures, adversary groups, malware, and their relationships. Each attack technique is related to data sources providing relevant information for its detection. 

We studied how many ATT\&CK techniques in the Enterprise ATT\&CK matrix of version 10.1 are visible on the network level, i.e., they have been assigned as data source one of: \textit{Network Traffic: Network Connection Creation}, \textit{Network Traffic: Network Traffic Content}, and \textit{Network Traffic: Network Traffic Flow}. We found that network traffic is relevant for 131 ATT\&CK techniques and sub-techniques from the overall amount of 707 techniques and sub-techniques. 

These techniques belong to 13 out of 14 phases of the ATT\&CK kill chain, i.e., 13 tactics. Therefore, MITRE ATT\&CK seems useful for modeling cyber threats when network monitoring is used as a data source. Such modeling may lead to methods operating on a sufficiently high level of maturity, as explained in Section \ref{sec:position}.

The overall analysis of data sources in ATT\&CK showed that the first place among all data sources belongs to \textit{Command} (256 techniques and sub-techniques) and \textit{Process} (253 techniques and sub-techniques), followed by \textit{File} (192) and \textit{Network Traffic} (131). There is a considerable gap between \textit{Network Traffic} and the fifth data source -- \textit{Windows Registry} (see Table \ref{tab:attack_sources}). Many techniques have no data source assigned.

\begin{table}[t]
    \centering
    \caption{The most common data sources in MITRE ATT\&CK (22nd March 2022).}
    \begin{tabular}{l|c}
        \toprule
        \textbf{Data Source} & \textbf{Count of Techniques} \\ \midrule
        Command & 256 \\
        Process & 253 \\
        File & 192 \\
        Network Traffic & 131 \\
        Windows Registry & 69 \\
        Application Log & 55 \\
        Module & 50 \\ \bottomrule
    \end{tabular}
    \label{tab:attack_sources}
\end{table}

We used references of enumerations denoted by arrows in Figure \ref{fig:refer_new} to answer the third research question. Each entry may have zero or more references to other entries. Consistent use of these references can allow expressing the CTI, e.g., using enumeration entries represented by several STIX objects. For example, a newly discovered vulnerability (CVE) can be exploited by several attack patterns (CAPEC, ATT\&CK). It would be beneficial to use CAPEC or ATT\&CK entries as detailed categories for vulnerability exploits.

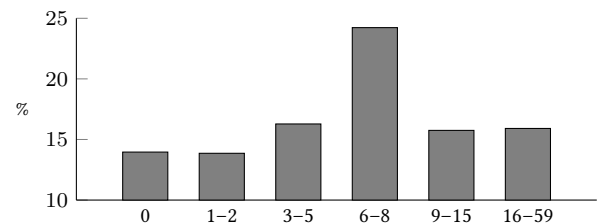
\begin{figure}[b]
    \centering
    \begin{tikzpicture}
    \begin{axis}[
        height = 4cm,
        width  = 8.5cm,
        symbolic x coords={0, 1--2, 3--5, 6--8, 9--15, 16--59},
        xtick=data,
        axis x line*=bottom,
        axis y line*=left,
        ylabel=\%,
        ylabel style={rotate=-90, font=\small},
        legend style={at={(0.5,-0.4)},
        anchor=north,legend columns=-1},
        ymin=10,
        ymax=25,
        x tick label style={font=\small},
        y tick label style={font=\small},
        bar width=0.6cm,
        enlarge x limits=0.18
      ]
        \addplot[ybar, fill=gray]
        coordinates {
            (0,   13.96)
            (1--2,  13.86)
            (3--5,   16.28)
            (6--8,   24.23)
            (9--15,  15.75)
            (16--59, 15.91)
        };
    \end{axis}
    \end{tikzpicture}
    \caption{How many CAPEC entries mapped to one CVE (15th January 2022).}
    \label{fig:cve_capec}
\end{figure}

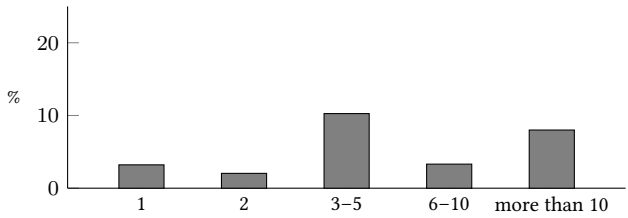
\begin{figure}[t]
    \centering
    \begin{tikzpicture}
    \begin{axis}[
        height = 4cm,
        width  = 9cm,
        symbolic x coords={0, 1, 2, 3--5, 6--10, more than 10},
        xtick=data,
        axis x line*=bottom,
        axis y line*=left,
        ylabel={\%},
        ylabel style={rotate=-90, font=\small},
        legend style={at={(0.5,-0.4)},
        anchor=north,legend columns=-1},
        ymin=0,
        ymax=25,
        x tick label style={font=\small},
        y tick label style={font=\small},
        bar width=0.6cm,
        enlarge x limits=0.18
      ]
        \addplot[ybar, fill=gray]
        coordinates {
            (0,   73.18)
            (1,  3.21)
            (2,   2.04)
            (3--5,  10.26)
            (6--10,  3.31)
            (more than 10, 8)
        };
    \end{axis}
    \end{tikzpicture}
    \caption{How many ATT\&CK techniques mapped to one CVE (15th January 2022).}
    \label{fig:cve_attack}
\end{figure}

Therefore, we took references of CVEs to their respective weaknesses (CWE). We complemented them with CWE entries' references to attack patterns (CAPEC). Finally, we added references of CAPEC entries to ATT\&CK techniques. Figure \ref{fig:cve_capec} shows that approximately 13\% of all vulnerabilities from the NVD do not have any related CAPEC. Then, 28.52\% of vulnerabilities have some meaningful count of CAPECs (one to five related CAPECs), and the remaining vulnerabilities have a higher count of CAPECs. As expected, more than 73\% of CVEs do not have any related ATT\&CK technique because CWE is application-oriented, and ATT\&CK is designated for network security (see Figure \ref{fig:cve_attack}). However, approximately a quarter of CVEs have some ATT\&CK technique, and more than 15\% have a meaningful count of them.

These results reveal that we cannot automatically determine which attack pattern or attack technique will be related to the vulnerability exploitation using references between enumerations in most cases. 
In other words, the CVE vulnerabilities from the NVD prefer such a subset of CWEs that does not allow mapping to a reasonable amount of CAPEC and ATT\&CK entries. 

A more detailed analysis of the references also revealed that only approximately one fourth of CWEs references some CAPECs. When CWE entry is related to one ATT\&CK technique, it usually has several CAPECs. It is mapped to only one ATT\&CK technique because only one out of several CAPECs is mapped to that ATT\&CK technique. Even if CWE transitively references one ATT\&CK technique, it does not necessarily lead to this ATT\&CK entry.

References between CAPEC and ATT\&CK can also be analyzed in the opposite way. Analysis of network-related ATT\&CK techniques at the beginning of this section revealed that only 48 out of these 131 techniques and sub-techniques have at least one assigned CAPEC. These ATT\&CK entries reference 22 CAPECs overall. Some ATT\&CK techniques are directly mapped to CAPECs, such as phishing. However, other techniques may have more than one CAPEC entry. Thus, references are not very useful for the transitive identification of threats in the organization's network.

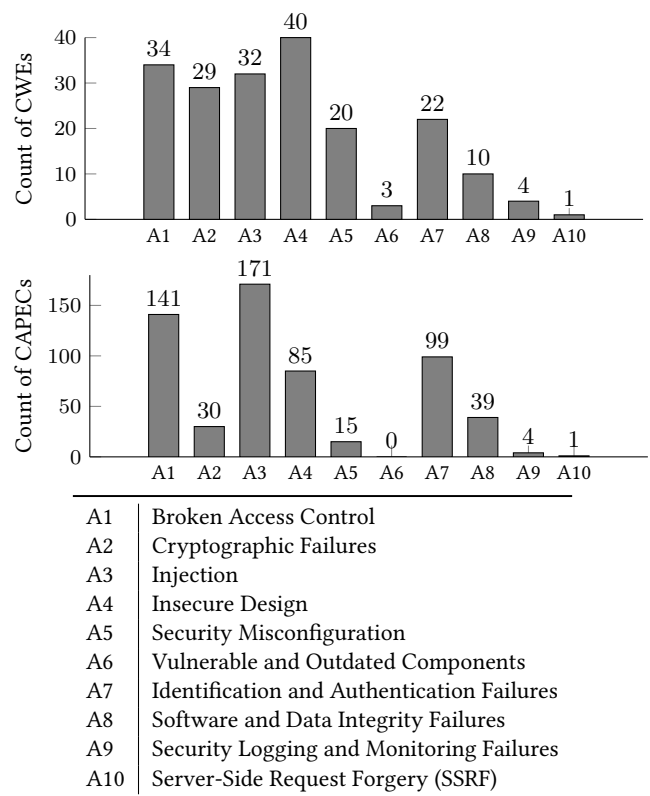
\begin{figure}
\centering
\begin{tikzpicture}
    \begin{axis}[
        height = 4cm,
        width  = 9cm,
        symbolic x coords={A1, A2, A3, A4, A5, A6, A7, A8, A9, A10},
        xtick=data,
        axis x line*=bottom,
        axis y line*=left,
        ylabel style={at={(-0.08, 0.5)}},
        legend style={at={(0.5,-0.4)},
        legend columns=1},
        nodes near coords,
        nodes near coords align={vertical},
        ylabel={Count of CWEs},
        ymin=0,
        ymax=40,
        x tick label style={font=\small},
        y tick label style={font=\small},
        bar width=0.4cm,
        enlarge x limits=0.18
      ]
        \addplot[ybar, fill=gray] 
        coordinates {
            (A1, 34)
            (A2, 29)
            (A3, 32)
            (A4, 40)
            (A5, 20)
            (A6, 3)
            (A7, 22)
            (A8, 10)
            (A9, 4)
            (A10, 1)
        };
    \end{axis}
\end{tikzpicture}

\begin{tikzpicture}
    \begin{axis}[
        height = 4cm,
        width  = 9cm,
        symbolic x coords={A1, A2, A3, A4, A5, A6, A7, A8, A9, A10},
        xtick=data,
        axis x line*=bottom,
        axis y line*=left,
        legend style={at={(0.5,-0.4)},
        legend columns=1},
        nodes near coords,
        nodes near coords align={vertical},
        ylabel={Count of CAPECs},
        ymin=0,
        ymax=180,
        x tick label style={font=\small},
        y tick label style={font=\small},
        bar width=0.4cm,
        enlarge x limits=0.18
      ]
        \addplot[ybar, fill=gray]
        coordinates {
            (A1, 141)
            (A2, 30)
            (A3, 171)
            (A4, 85)
            (A5, 15)
            (A6, 0)
            (A7, 99)
            (A8, 39)
            (A9, 4)
            (A10, 1)
        };
    \end{axis}
\end{tikzpicture}

\smallskip

\begin{tabular}{l|l}
    \toprule
    A1 & Broken Access Control \\
	A2 & Cryptographic Failures \\
	A3 & Injection \\
	A4 & Insecure Design \\
	A5 & Security Misconfiguration \\
    A6 & Vulnerable and Outdated Components \\
    A7 & Identification and Authentication Failures\\
    A8 & Software and Data Integrity Failures \\
    A9 & Security Logging and Monitoring Failures \\
    A10 & Server-Side Request Forgery (SSRF)\\
	\bottomrule
\end{tabular}
\label{fig:owasp_cwe_capec}
\caption{Count of CWEs and CAPECs for OWASP Top Ten 2021 (15th January 2022).}
\label{owasp_categories}
\end{figure}

We also studied interoperability with OWASP Top Ten list because results of scanning tools are often its categories. MITRE provides a mapping labeled as \textit{CWE-1344} that assigns to each category its related CWEs. Appropriate counts of CWEs and CAPECs for each category are depicted in Figure \ref{owasp_categories}. These counts are caused by the fact that CWEs are more granular compared to OWASP Top Ten. For example, OWASP Injection is related to SQL injection, OS command injection, and other types of injections in CWE. The categories have uneven amount of CAPECs too.

Based on these results, we conclude that mapping of enumeration entries may not accurately determine related parts of the information based on enumeration entries observed in the organization.
References between enumerations from Figure \ref{fig:refer_new} are ambiguous. The enumerations often contain free-text fields that enforce text-based approaches for their processing (e.g., using NLP). The text-based approach using Term Frequency -- Inverse Document Frequency (TF-IDF) together with graph-based and recommendation-based approaches were provided by \textit{Dang and Fran\c{c}ois}~\cite{Dang2018}.

\section{Conclusion} \label{sec:conclusion}
In this paper, we surveyed cyber threat and vulnerability identification using enumerations about vulnerabilities and attack techniques. We revealed the current challenges and possible research directions suitable for future work. Last, we analyzed data from MITRE ATT\&CK and enumerations using data sources of ATT\&CK techniques and interoperability of enumeration using references between their entries.

In general, current vulnerability identification methods can determine CPE and CVE identifiers in practice. The challenges are related to the vulnerability discovery precision, a large amount of data, hybrid IT environments, and conformance to specifications. The threat identification copes, e.g., with the unstructured text of CTI reports, the overwhelming amount of data, and related lack of visibility. Enumeration entries allow the development of mature methods based on the Detection Maturity Level (DML) model.

According to the analytical results, ATT\&CK provides capabilities for complex threat modeling when only network monitoring is used as a source of data. However, references between enumerations may not allow inferring related parts of the information about possible cyber threats in the organization. Scripts that allow reproducing results are provided in supplementary materials in the ACM Digital Library and on Zenodo \cite{supplementary}.

\begin{acks}
This research was supported by the CONCORDIA project that has received funding from the European Union’s Horizon 2020 research and innovation programme under the grant agreement No 830927.
\end{acks}

\balance
\bibliographystyle{ACM-Reference-Format}
\bibliography{bibliography}


\begin{thebibliography}{53}


\ifx \showCODEN    \undefined \def \showCODEN     #1{\unskip}     \fi
\ifx \showDOI      \undefined \def \showDOI       #1{#1}\fi
\ifx \showISBNx    \undefined \def \showISBNx     #1{\unskip}     \fi
\ifx \showISBNxiii \undefined \def \showISBNxiii  #1{\unskip}     \fi
\ifx \showISSN     \undefined \def \showISSN      #1{\unskip}     \fi
\ifx \showLCCN     \undefined \def \showLCCN      #1{\unskip}     \fi
\ifx \shownote     \undefined \def \shownote      #1{#1}          \fi
\ifx \showarticletitle \undefined \def \showarticletitle #1{#1}   \fi
\ifx \showURL      \undefined \def \showURL       {\relax}        \fi
\providecommand\bibfield[2]{#2}
\providecommand\bibinfo[2]{#2}
\providecommand\natexlab[1]{#1}
\providecommand\showeprint[2][]{arXiv:#2}

\bibitem[Bianco(2013)]%
        {bianco2013}
\bibfield{author}{\bibinfo{person}{David Bianco}.}
  \bibinfo{year}{2013}\natexlab{}.
\newblock \bibinfo{title}{{The Pyramid of Pain}}.
\newblock
\newblock
\urldef\tempurl%
\url{http://detect-respond.blogspot.com/2013/03/the-pyramid-of-pain.html}
\showURL{%
Retrieved Jan 29, 2022 from \tempurl}


\bibitem[Bodeau et~al\mbox{.}(2018)]%
        {Bodeau2018}
\bibfield{author}{\bibinfo{person}{Deborah~J. Bodeau},
  \bibinfo{person}{Catherine~D. McCollum}, {and} \bibinfo{person}{David~B.
  Fox}.} \bibinfo{year}{2018}\natexlab{}.
\newblock \bibinfo{booktitle}{\emph{{Cyber Threat Modeling: Survey, Assessment,
  and Representative Framework}}}.
\newblock \bibinfo{type}{{T}echnical {R}eport}. \bibinfo{institution}{The MITRE
  Corporation}, \bibinfo{address}{McLean, Virginia}. \bibinfo{pages}{1--111}
  pages.
\newblock
\urldef\tempurl%
\url{https://apps.dtic.mil/sti/pdfs/AD1108051.pdf}
\showURL{%
Retrieved Jan 29, 2022 from \tempurl}


\bibitem[Bou-Harb et~al\mbox{.}(2014)]%
        {BouHarb2014}
\bibfield{author}{\bibinfo{person}{Elias Bou-Harb}, \bibinfo{person}{Mourad
  Debbabi}, {and} \bibinfo{person}{Chadi Assi}.}
  \bibinfo{year}{2014}\natexlab{}.
\newblock \showarticletitle{{Cyber scanning: A comprehensive survey}}.
\newblock \bibinfo{journal}{\emph{IEEE Communications Surveys and Tutorials}}
  \bibinfo{volume}{16} (\bibinfo{year}{2014}), \bibinfo{pages}{1496--1519}.
\newblock
Issue 3.
\showISSN{1553877X}
\urldef\tempurl%
\url{https://doi.org/10.1109/SURV.2013.102913.00020}
\showDOI{\tempurl}


\bibitem[Brown and Lee(2019)]%
        {brown2019evolution}
\bibfield{author}{\bibinfo{person}{Rebekah Brown} {and}
  \bibinfo{person}{Robert~M. Lee}.} \bibinfo{year}{2019}\natexlab{}.
\newblock \bibinfo{booktitle}{\emph{{The Evolution of Cyber Threat Intelligence
  (CTI): 2019 SANS CTI Survey}}}.
\newblock SANS Institute.
\newblock
\urldef\tempurl%
\url{https://www.sans.org/white-papers/38790/}
\showURL{%
Retrieved Jan 29, 2022 from \tempurl}


\bibitem[Cheikes et~al\mbox{.}(2011)]%
        {cpe}
\bibfield{author}{\bibinfo{person}{Brant~A. Cheikes}, \bibinfo{person}{David
  Waltermire}, {and} \bibinfo{person}{Karen Scarfone}.}
  \bibinfo{year}{2011}\natexlab{}.
\newblock \bibinfo{booktitle}{\emph{{Common platform enumeration: Naming
  specification version 2.3}}}.
\newblock National Institute of Standards and Technology.
\newblock
\urldef\tempurl%
\url{https://nvlpubs.nist.gov/nistpubs/Legacy/IR/nistir7695.pdf}
\showURL{%
Retrieved Jan 29, 2022 from \tempurl}


\bibitem[Corporation(2022b)]%
        {cve}
\bibfield{author}{\bibinfo{person}{The~MITRE Corporation}.}
  \bibinfo{year}{1999-2022}\natexlab{b}.
\newblock \bibinfo{title}{{CVE - Common Vulnerabilities and Exposures (CVE)}}.
\newblock
\newblock
\urldef\tempurl%
\url{https://cve.mitre.org/}
\showURL{%
Retrieved Jan 29, 2022 from \tempurl}


\bibitem[Corporation(2022c)]%
        {cwe}
\bibfield{author}{\bibinfo{person}{The~MITRE Corporation}.}
  \bibinfo{year}{2006-2022}\natexlab{c}.
\newblock \bibinfo{title}{{CWE - Common Weakness Enumeration}}.
\newblock
\newblock
\urldef\tempurl%
\url{https://cwe.mitre.org/}
\showURL{%
Retrieved Jan 29, 2022 from \tempurl}


\bibitem[Corporation(2022a)]%
        {capec}
\bibfield{author}{\bibinfo{person}{The~MITRE Corporation}.}
  \bibinfo{year}{2007-2022}\natexlab{a}.
\newblock \bibinfo{title}{{CAPEC - Common Attack Pattern Enumeration and
  Classification (CAPEC)}}.
\newblock
\newblock
\urldef\tempurl%
\url{https://capec.mitre.org/}
\showURL{%
Retrieved Jan 29, 2022 from \tempurl}


\bibitem[Corporation(2022e)]%
        {attack}
\bibfield{author}{\bibinfo{person}{The~MITRE Corporation}.}
  \bibinfo{year}{2015-2022}\natexlab{e}.
\newblock \bibinfo{title}{{MITRE ATT\&CK}}.
\newblock
\newblock
\urldef\tempurl%
\url{https://attack.mitre.org/}
\showURL{%
Retrieved Jan 29, 2022 from \tempurl}


\bibitem[Corporation(2022d)]%
        {mitrecti}
\bibfield{author}{\bibinfo{person}{The~MITRE Corporation}.}
  \bibinfo{year}{2022}\natexlab{d}.
\newblock \bibinfo{title}{{Github - mitre/cti: Cyber Threat Intelligence
  Repository expressed in STIX 2.0}}.
\newblock
\newblock
\urldef\tempurl%
\url{https://github.com/mitre/cti}
\showURL{%
Retrieved Mar 22, 2022 from \tempurl}


\bibitem[Cybersecurity(2022)]%
        {alienvaultusm}
\bibfield{author}{\bibinfo{person}{AT\&T Cybersecurity}.}
  \bibinfo{year}{2022}\natexlab{}.
\newblock \bibinfo{title}{{Threat Detection \& Response}}.
\newblock
\newblock
\urldef\tempurl%
\url{https://cybersecurity.att.com/solutions/threat-detection}
\showURL{%
Retrieved Feb 1, 2022 from \tempurl}


\bibitem[Dandurand et~al\mbox{.}(2014)]%
        {pawlinski2014standards}
\bibfield{author}{\bibinfo{person}{Luc Dandurand}, \bibinfo{person}{Aaron
  Kaplan}, \bibinfo{person}{Pavel K{\'a}cha}, \bibinfo{person}{Youki
  Kadobayashi}, \bibinfo{person}{Andrew Kompanek}, \bibinfo{person}{Tom{\'a}s
  Lima}, \bibinfo{person}{Thomas Millar}, \bibinfo{person}{Jose Nazario},
  \bibinfo{person}{Richard Perlotto}, {and} \bibinfo{person}{Wes Young}.}
  \bibinfo{year}{2014}\natexlab{}.
\newblock \bibinfo{booktitle}{\emph{{Standards and tools for exchange and
  processing of actionable information}}}.
\newblock \bibinfo{type}{{T}echnical {R}eport}. \bibinfo{institution}{European
  Union Agency for Network and Information Security (ENISA)},
  \bibinfo{address}{Heraklion, Greece}.
\newblock
\showISBNx{978-92-9204-105-2}
\urldef\tempurl%
\url{https://doi.org/10.2824/37776}
\showDOI{\tempurl}


\bibitem[Dang and Fran\c{c}ois(2018)]%
        {Dang2018}
\bibfield{author}{\bibinfo{person}{Quang~Vinh Dang} {and}
  \bibinfo{person}{Jerome Fran\c{c}ois}.} \bibinfo{year}{2018}\natexlab{}.
\newblock \showarticletitle{{Utilizing attack enumerations to study SDN/NFV
  vulnerabilities}}. In \bibinfo{booktitle}{\emph{2018 4th IEEE Conference on
  Network Softwarization and Workshops, NetSoft 2018}}.
  \bibinfo{publisher}{IEEE}, \bibinfo{address}{New York, NY, USA},
  \bibinfo{pages}{329--331}.
\newblock
\showISBNx{9781538646335}
\urldef\tempurl%
\url{https://doi.org/10.1109/NETSOFT.2018.8459961}
\showDOI{\tempurl}


\bibitem[Edgescan(2020)]%
        {Edgescan2020}
\bibfield{author}{\bibinfo{person}{Edgescan}.} \bibinfo{year}{2020}\natexlab{}.
\newblock \bibinfo{title}{{The Year 2020 Vulnerability statistics report}}.
\newblock
\newblock
\urldef\tempurl%
\url{https://info.edgescan.com/vulnerability-stats}
\showURL{%
Retrieved Jan 29, 2022 from \tempurl}


\bibitem[{Gartner, Inc}(2022)]%
        {gartnersoar}
\bibfield{author}{\bibinfo{person}{{Gartner, Inc}}.}
  \bibinfo{year}{2022}\natexlab{}.
\newblock \bibinfo{title}{{Security Orchestration, Automation and Response
  (SOAR)}}.
\newblock
\newblock
\urldef\tempurl%
\url{https://www.gartner.com/en/information-technology/glossary/security-orchestration-automation-response-soar}
\showURL{%
Retrieved May 21, 2022 from \tempurl}


\bibitem[Gawron et~al\mbox{.}(2017)]%
        {Gawron2017}
\bibfield{author}{\bibinfo{person}{Marian Gawron}, \bibinfo{person}{Feng
  Cheng}, {and} \bibinfo{person}{Christoph Meinel}.}
  \bibinfo{year}{2017}\natexlab{}.
\newblock \showarticletitle{{PVD: Passive vulnerability detection}}. In
  \bibinfo{booktitle}{\emph{2017 8th International Conference on Information
  and Communication Systems (ICICS)}}. \bibinfo{publisher}{IEEE},
  \bibinfo{address}{New York, NY, USA}, \bibinfo{pages}{322--327}.
\newblock
\showISBNx{978-1-5090-4243-2}
\urldef\tempurl%
\url{https://doi.org/10.1109/IACS.2017.7921992}
\showDOI{\tempurl}


\bibitem[Genge and Enăchescu(2016)]%
        {Genge2016}
\bibfield{author}{\bibinfo{person}{Béla Genge} {and} \bibinfo{person}{Călin
  Enăchescu}.} \bibinfo{year}{2016}\natexlab{}.
\newblock \showarticletitle{{ShoVAT: Shodan-based vulnerability assessment tool
  for Internet-facing services}}.
\newblock \bibinfo{journal}{\emph{Security and Communication Networks}}
  \bibinfo{volume}{9}, \bibinfo{number}{15} (\bibinfo{date}{10}
  \bibinfo{year}{2016}), \bibinfo{pages}{2696--2714}.
\newblock
\showISSN{19390122}
\urldef\tempurl%
\url{https://doi.org/10.1002/sec.1262}
\showDOI{\tempurl}


\bibitem[Hofstede et~al\mbox{.}(2014)]%
        {rick2014flow}
\bibfield{author}{\bibinfo{person}{Rick Hofstede}, \bibinfo{person}{Pavel
  Čeleda}, \bibinfo{person}{Brian Trammell}, \bibinfo{person}{Idilio Drago},
  \bibinfo{person}{Ramin Sadre}, \bibinfo{person}{Anna Sperotto}, {and}
  \bibinfo{person}{Aiko Pras}.} \bibinfo{year}{2014}\natexlab{}.
\newblock \showarticletitle{Flow Monitoring Explained: From Packet Capture to
  Data Analysis With NetFlow and IPFIX}.
\newblock \bibinfo{journal}{\emph{IEEE Communications Surveys and Tutorials}}
  \bibinfo{volume}{16}, \bibinfo{number}{4} (\bibinfo{year}{2014}),
  \bibinfo{pages}{2037--2064}.
\newblock
\urldef\tempurl%
\url{https://doi.org/10.1109/COMST.2014.2321898}
\showDOI{\tempurl}


\bibitem[Husari et~al\mbox{.}(2017)]%
        {Husari2017}
\bibfield{author}{\bibinfo{person}{Ghaith Husari}, \bibinfo{person}{Ehab
  Al-Shaer}, \bibinfo{person}{Mohiuddin Ahmed}, \bibinfo{person}{Bill Chu},
  {and} \bibinfo{person}{Xi Niu}.} \bibinfo{year}{2017}\natexlab{}.
\newblock \showarticletitle{{TTPDrill: Automatic and accurate extraction of
  threat actions from unstructured text of CTI Sources}}. In
  \bibinfo{booktitle}{\emph{Proceedings of the 33rd Annual Computer Security
  Applications Conference}} \emph{(\bibinfo{series}{ACSAC 2017})}.
  \bibinfo{publisher}{ACM}, \bibinfo{address}{New York, NY, USA},
  \bibinfo{pages}{103--115}.
\newblock
\showISBNx{9781450353458}
\urldef\tempurl%
\url{https://doi.org/10.1145/3134600.3134646}
\showDOI{\tempurl}


\bibitem[Hutchins et~al\mbox{.}(2011)]%
        {hutchins2011}
\bibfield{author}{\bibinfo{person}{Eric~M. Hutchins},
  \bibinfo{person}{Michael~J. Cloppert}, {and} \bibinfo{person}{Rohan~M.
  Amin}.} \bibinfo{year}{2011}\natexlab{}.
\newblock \showarticletitle{{Intelligence-driven computer network defense
  informed by analysis of adversary campaigns and intrusion kill chains}}.
\newblock \bibinfo{journal}{\emph{Leading Issues in Information Warfare \&
  Security Research}} \bibinfo{volume}{1}, \bibinfo{number}{1}
  (\bibinfo{year}{2011}), \bibinfo{pages}{80--106}.
\newblock


\bibitem[IBM(2022)]%
        {IBMQRadar}
\bibfield{author}{\bibinfo{person}{IBM}.} \bibinfo{year}{2022}\natexlab{}.
\newblock \bibinfo{title}{{IBM Security QRadar SIEM - Features}}.
\newblock
\newblock
\urldef\tempurl%
\url{https://www.ibm.com/qradar/security-qradar-siem/features}
\showURL{%
Retrieved Feb 2, 2022 from \tempurl}


\bibitem[Kaloudi and Li(2020)]%
        {Kaloudi2020ai}
\bibfield{author}{\bibinfo{person}{Nektaria Kaloudi} {and}
  \bibinfo{person}{Jingyue Li}.} \bibinfo{year}{2020}\natexlab{}.
\newblock \showarticletitle{The AI-Based Cyber Threat Landscape: A Survey}.
\newblock \bibinfo{journal}{\emph{ACM Computing Surveys (CSUR)}}
  \bibinfo{volume}{53}, \bibinfo{number}{1} (\bibinfo{year}{2020}),
  \bibinfo{pages}{1--34}.
\newblock
\urldef\tempurl%
\url{https://doi.org/10.1145/3372823}
\showDOI{\tempurl}


\bibitem[Laštovička et~al\mbox{.}(2020)]%
        {Lastovicka2020}
\bibfield{author}{\bibinfo{person}{Martin Laštovička},
  \bibinfo{person}{Martin Husák}, {and} \bibinfo{person}{Lukáš Sadlek}.}
  \bibinfo{year}{2020}\natexlab{}.
\newblock \showarticletitle{{Network Monitoring and Enumerating Vulnerabilities
  in Large Heterogeneous Networks}}. In \bibinfo{booktitle}{\emph{NOMS 2020 -
  2020 IEEE/IFIP Network Operations and Management Symposium}}.
  \bibinfo{publisher}{IEEE}, \bibinfo{address}{New York, NY, USA},
  \bibinfo{pages}{1--6}.
\newblock
\showISBNx{978-1-7281-4973-8}
\urldef\tempurl%
\url{https://doi.org/10.1109/NOMS47738.2020.9110394}
\showDOI{\tempurl}


\bibitem[Le et~al\mbox{.}(2021)]%
        {le2021survey}
\bibfield{author}{\bibinfo{person}{Triet H.~M. Le}, \bibinfo{person}{Huaming
  Chen}, {and} \bibinfo{person}{M.~Ali Babar}.}
  \bibinfo{year}{2021}\natexlab{}.
\newblock \bibinfo{title}{A Survey on Data-driven Software Vulnerability
  Assessment and Prioritization}.
\newblock \bibinfo{howpublished}{arXiv}.
\newblock
\urldef\tempurl%
\url{https://doi.org/10.48550/ARXIV.2107.08364}
\showDOI{\tempurl}


\bibitem[Lyon(2009)]%
        {nmap}
\bibfield{author}{\bibinfo{person}{Gordon~Fyodor Lyon}.}
  \bibinfo{year}{2009}\natexlab{}.
\newblock \bibinfo{booktitle}{\emph{{Nmap Network Scanning: The Official Nmap
  Project Guide to Network Discovery and Security Scanning}}}.
\newblock \bibinfo{publisher}{Insecure}, \bibinfo{address}{USA}.
\newblock
\showISBNx{0979958717, 9780979958717}
\urldef\tempurl%
\url{https://nmap.org/book/}
\showURL{%
\tempurl}


\bibitem[Mavroeidis and Bromander(2017)]%
        {Mavroeidis2017}
\bibfield{author}{\bibinfo{person}{Vasileios Mavroeidis} {and}
  \bibinfo{person}{Siri Bromander}.} \bibinfo{year}{2017}\natexlab{}.
\newblock \showarticletitle{{Cyber Threat Intelligence Model: An Evaluation of
  Taxonomies, Sharing Standards, and Ontologies within Cyber Threat
  Intelligence}}. In \bibinfo{booktitle}{\emph{2017 European Intelligence and
  Security Informatics Conference (EISIC)}}. \bibinfo{publisher}{IEEE},
  \bibinfo{address}{New York, NY, USA}, \bibinfo{pages}{91--98}.
\newblock
\showISBNx{978-1-5386-2385-5}
\urldef\tempurl%
\url{https://doi.org/10.1109/EISIC.2017.20}
\showDOI{\tempurl}


\bibitem[Mavroeidis and Jøsang(2018)]%
        {Mavroeidis2018}
\bibfield{author}{\bibinfo{person}{Vasileios Mavroeidis} {and}
  \bibinfo{person}{Audun Jøsang}.} \bibinfo{year}{2018}\natexlab{}.
\newblock \showarticletitle{{Data-Driven Threat Hunting Using Sysmon}}. In
  \bibinfo{booktitle}{\emph{Proceedings of the 2nd International Conference on
  Cryptography, Security and Privacy}} \emph{(\bibinfo{series}{ICCSP 2018})}.
  \bibinfo{publisher}{ACM}, \bibinfo{address}{New York, NY, USA},
  \bibinfo{pages}{82–88}.
\newblock
\showISBNx{9781450363617}
\urldef\tempurl%
\url{https://doi.org/10.1145/3199478.3199490}
\showDOI{\tempurl}


\bibitem[Milajerdi et~al\mbox{.}(2019)]%
        {Milajerdi2019Holmes}
\bibfield{author}{\bibinfo{person}{Sadegh~M. Milajerdi}, \bibinfo{person}{Rigel
  Gjomemo}, \bibinfo{person}{Birhanu Eshete}, \bibinfo{person}{R. Sekar}, {and}
  \bibinfo{person}{V.~N. Venkatakrishnan}.} \bibinfo{year}{2019}\natexlab{}.
\newblock \showarticletitle{{HOLMES: Real-Time APT Detection through
  Correlation of Suspicious Information Flows}}. In
  \bibinfo{booktitle}{\emph{2019 IEEE Symposium on Security and Privacy (SP)}}.
  \bibinfo{publisher}{IEEE}, \bibinfo{address}{New York, NY, USA},
  \bibinfo{pages}{1137--1152}.
\newblock
\showISBNx{978-1-5386-6660-9}
\urldef\tempurl%
\url{https://doi.org/10.1109/SP.2019.00026}
\showDOI{\tempurl}


\bibitem[Muniz et~al\mbox{.}(2016)]%
        {Muniz2016}
\bibfield{author}{\bibinfo{person}{Joseph Muniz}, \bibinfo{person}{Gary
  McIntyre}, {and} \bibinfo{person}{Nadhem AlFardan}.}
  \bibinfo{year}{2016}\natexlab{}.
\newblock \bibinfo{booktitle}{\emph{Security Operations Center: Building,
  Operating, and Maintaining Your SOC}}.
\newblock \bibinfo{publisher}{Cisco Press}, \bibinfo{address}{Indianapolis
  (USA)}.
\newblock
\showISBNx{978-0-13-405201-4}


\bibitem[Noor et~al\mbox{.}(2019)]%
        {Noor2019}
\bibfield{author}{\bibinfo{person}{Umara Noor}, \bibinfo{person}{Zahid Anwar},
  \bibinfo{person}{Asad~Waqar Malik}, \bibinfo{person}{Sharifullah Khan}, {and}
  \bibinfo{person}{Shahzad Saleem}.} \bibinfo{year}{2019}\natexlab{}.
\newblock \showarticletitle{{A machine learning framework for investigating
  data breaches based on semantic analysis of adversary’s attack patterns in
  threat intelligence repositories}}.
\newblock \bibinfo{journal}{\emph{Future Generation Computer Systems}}
  \bibinfo{volume}{95} (\bibinfo{date}{6} \bibinfo{year}{2019}),
  \bibinfo{pages}{467--487}.
\newblock
\showISSN{0167739X}
\urldef\tempurl%
\url{https://doi.org/10.1016/j.future.2019.01.022}
\showDOI{\tempurl}


\bibitem[of~Standards and (NIST)(2018)]%
        {nistcyberframework}
\bibfield{author}{\bibinfo{person}{National~Institute of Standards} {and}
  \bibinfo{person}{Technology (NIST)}.} \bibinfo{year}{2018}\natexlab{}.
\newblock \bibinfo{title}{{Framework for Improving Critical Infrastructure
  Cybersecurity}}.
\newblock
\newblock
\urldef\tempurl%
\url{https://nvlpubs.nist.gov/nistpubs/CSWP/NIST.CSWP.04162018.pdf}
\showURL{%
Retrieved Jan 29, 2022 from \tempurl}


\bibitem[of~Standards and Technology(2022a)]%
        {nvd}
\bibfield{author}{\bibinfo{person}{The National~Institute of Standards} {and}
  \bibinfo{person}{Technology}.} \bibinfo{year}{2022}\natexlab{a}.
\newblock \bibinfo{title}{{NVD - General}}.
\newblock
\newblock
\urldef\tempurl%
\url{https://nvd.nist.gov/general}
\showURL{%
Retrieved May 19, 2022 from \tempurl}


\bibitem[of~Standards and Technology(2022b)]%
        {cpe_statistics}
\bibfield{author}{\bibinfo{person}{The National~Institute of Standards} {and}
  \bibinfo{person}{Technology}.} \bibinfo{year}{2022}\natexlab{b}.
\newblock \bibinfo{title}{{Official Common Platform Enumeration (CPE)
  Dictionary Statistics}}.
\newblock
\newblock
\urldef\tempurl%
\url{https://nvd.nist.gov/products/cpe/statistics}
\showURL{%
Retrieved Jan 29, 2022 from \tempurl}


\bibitem[O'Hare et~al\mbox{.}(2019)]%
        {Hare2019}
\bibfield{author}{\bibinfo{person}{Jamie O'Hare}, \bibinfo{person}{Rich
  Macfarlane}, {and} \bibinfo{person}{Owen Lo}.}
  \bibinfo{year}{2019}\natexlab{}.
\newblock \showarticletitle{{Identifying Vulnerabilities Using Internet-Wide
  Scanning Data}}. In \bibinfo{booktitle}{\emph{2019 IEEE 12th International
  Conference on Global Security, Safety and Sustainability (ICGS3)}}.
  \bibinfo{publisher}{IEEE}, \bibinfo{address}{New York, NY, USA},
  \bibinfo{pages}{1--10}.
\newblock
\showISBNx{978-1-5386-7001-9}
\urldef\tempurl%
\url{https://doi.org/10.1109/ICGS3.2019.8688018}
\showDOI{\tempurl}


\bibitem[Piazza et~al\mbox{.}(2017)]%
        {stixpart1}
\bibfield{author}{\bibinfo{person}{Rich Piazza}, \bibinfo{person}{John Wunder},
  {and} \bibinfo{person}{Bret Jordan}.} \bibinfo{year}{2017}\natexlab{}.
\newblock \bibinfo{title}{{STIX™ Version 2.0. Part 1: STIX Core Concepts}}.
\newblock
\newblock
\urldef\tempurl%
\url{http://docs.oasis-open.org/cti/stix/v2.0/stix-v2.0-part1-stix-core.html}
\showURL{%
Retrieved Jan 29, 2022 from \tempurl}


\bibitem[{Ponemon Institute, LLC.}(2019)]%
        {ponemon2019}
\bibfield{author}{\bibinfo{person}{{Ponemon Institute, LLC.}}}
  \bibinfo{year}{2019}\natexlab{}.
\newblock \bibinfo{title}{{Costs and consequences of gaps in vulnerability
  response}}.
\newblock
\newblock
\urldef\tempurl%
\url{https://www.servicenow.com/lpayr/ponemon-vulnerability-survey.html}
\showURL{%
Retrieved Jan 29, 2022 from \tempurl}


\bibitem[Qamar et~al\mbox{.}(2017)]%
        {Qamar2017}
\bibfield{author}{\bibinfo{person}{Sara Qamar}, \bibinfo{person}{Zahid Anwar},
  \bibinfo{person}{Mohammad~Ashiqur Rahman}, \bibinfo{person}{Ehab Al-Shaer},
  {and} \bibinfo{person}{Bei-Tseng Chu}.} \bibinfo{year}{2017}\natexlab{}.
\newblock \showarticletitle{{Data-driven analytics for cyber-threat
  intelligence and information sharing}}.
\newblock \bibinfo{journal}{\emph{Computers \& Security}}  \bibinfo{volume}{67}
  (\bibinfo{date}{6} \bibinfo{year}{2017}), \bibinfo{pages}{35--58}.
\newblock
\showISSN{01674048}
\urldef\tempurl%
\url{https://doi.org/10.1016/j.cose.2017.02.005}
\showDOI{\tempurl}


\bibitem[{Qualys, Inc.}(2022)]%
        {QualysVM}
\bibfield{author}{\bibinfo{person}{{Qualys, Inc.}}}
  \bibinfo{year}{2022}\natexlab{}.
\newblock \bibinfo{title}{{Vulnerability Management}}.
\newblock
\newblock
\urldef\tempurl%
\url{https://www.qualys.com/apps/vulnerability-management/}
\showURL{%
Retrieved Feb 2, 2022 from \tempurl}


\bibitem[Rahman et~al\mbox{.}(2021)]%
        {rahman2021attackers}
\bibfield{author}{\bibinfo{person}{Md~Rayhanur Rahman}, \bibinfo{person}{Rezvan
  Mahdavi-Hezaveh}, {and} \bibinfo{person}{Laurie Williams}.}
  \bibinfo{year}{2021}\natexlab{}.
\newblock \bibinfo{title}{What are the attackers doing now? Automating cyber
  threat intelligence extraction from text on pace with the changing threat
  landscape: A survey}.
\newblock
\newblock
\urldef\tempurl%
\url{https://doi.org/10.48550/ARXIV.2109.06808}
\showDOI{\tempurl}


\bibitem[Rapid7(2022)]%
        {metasploit}
\bibfield{author}{\bibinfo{person}{Rapid7}.} \bibinfo{year}{2022}\natexlab{}.
\newblock \bibinfo{title}{{Metasploit Documentation}}.
\newblock
\newblock
\urldef\tempurl%
\url{https://docs.rapid7.com/metasploit/}
\showURL{%
Retrieved Jan 28, 2022 from \tempurl}


\bibitem[Sadlek et~al\mbox{.}(2022)]%
        {supplementary}
\bibfield{author}{\bibinfo{person}{Luk\'{a}\v{s} Sadlek},
  \bibinfo{person}{Pavel \v{C}eleda}, {and} \bibinfo{person}{Daniel
  Tovar\v{n}\'{a}k}.} \bibinfo{year}{2022}\natexlab{}.
\newblock \bibinfo{booktitle}{\emph{{Supplementary Materials: Current
  Challenges of Cyber Threat and Vulnerability Identification Using Public
  Enumerations}}}.
\newblock Zenodo.
\newblock
\urldef\tempurl%
\url{https://doi.org/10.5281/zenodo.6659657}
\showURL{%
Retrieved Jun 20, 2022 from \tempurl}


\bibitem[Samtani et~al\mbox{.}(2016)]%
        {Samtani2016}
\bibfield{author}{\bibinfo{person}{Sagar Samtani}, \bibinfo{person}{Shuo Yu},
  \bibinfo{person}{Hongyi Zhu}, \bibinfo{person}{Mark Patton}, {and}
  \bibinfo{person}{Hsinchun Chen}.} \bibinfo{year}{2016}\natexlab{}.
\newblock \showarticletitle{{Identifying SCADA vulnerabilities using passive
  and active vulnerability assessment techniques}}. In
  \bibinfo{booktitle}{\emph{2016 IEEE Conference on Intelligence and Security
  Informatics (ISI)}}. \bibinfo{publisher}{IEEE}, \bibinfo{address}{New York,
  NY, USA}, \bibinfo{pages}{25--30}.
\newblock
\showISBNx{978-1-5090-3865-7}
\urldef\tempurl%
\url{https://doi.org/10.1109/ISI.2016.7745438}
\showDOI{\tempurl}


\bibitem[Sauerwein et~al\mbox{.}(2019)]%
        {Sauerwein2019}
\bibfield{author}{\bibinfo{person}{Clemens Sauerwein}, \bibinfo{person}{Irdin
  Pekaric}, \bibinfo{person}{Michael Felderer}, {and} \bibinfo{person}{Ruth
  Breu}.} \bibinfo{year}{2019}\natexlab{}.
\newblock \showarticletitle{{An analysis and classification of public
  information security data sources used in research and practice}}.
\newblock \bibinfo{journal}{\emph{Computers \& Security}}  \bibinfo{volume}{82}
  (\bibinfo{date}{5} \bibinfo{year}{2019}), \bibinfo{pages}{140--155}.
\newblock
\showISSN{01674048}
\urldef\tempurl%
\url{https://doi.org/10.1016/j.cose.2018.12.011}
\showDOI{\tempurl}


\bibitem[Security(2022)]%
        {exploitdb}
\bibfield{author}{\bibinfo{person}{Offensive Security}.}
  \bibinfo{year}{2022}\natexlab{}.
\newblock \bibinfo{title}{{Exploit Database}}.
\newblock
\newblock
\urldef\tempurl%
\url{https://www.exploit-db.com/}
\showURL{%
Retrieved Jan 28, 2022 from \tempurl}


\bibitem[Stillions(2014)]%
        {stillions2014}
\bibfield{author}{\bibinfo{person}{Ryan Stillions}.}
  \bibinfo{year}{2014}\natexlab{}.
\newblock \bibinfo{title}{{The DML model}}.
\newblock
\newblock
\urldef\tempurl%
\url{http://ryanstillions.blogspot.com/2014/04/the-dml-model_21.html}
\showURL{%
Retrieved Jan 29, 2022 from \tempurl}


\bibitem[{Tenable, Inc.}(2022a)]%
        {nessus}
\bibfield{author}{\bibinfo{person}{{Tenable, Inc.}}}
  \bibinfo{year}{2022}\natexlab{a}.
\newblock \bibinfo{title}{{Nessus}}.
\newblock
\newblock
\urldef\tempurl%
\url{https://www.tenable.com/products/nessus}
\showURL{%
Retrieved Jan 28, 2022 from \tempurl}


\bibitem[{Tenable, Inc.}(2022b)]%
        {TenableVM}
\bibfield{author}{\bibinfo{person}{{Tenable, Inc.}}}
  \bibinfo{year}{2022}\natexlab{b}.
\newblock \bibinfo{title}{{Tenable.io vulnerability management}}.
\newblock
\newblock
\urldef\tempurl%
\url{https://www.tenable.com/products/tenable-io}
\showURL{%
Retrieved Feb 2, 2022 from \tempurl}


\bibitem[Tounsi and Rais(2018)]%
        {Tounsi2018}
\bibfield{author}{\bibinfo{person}{Wiem Tounsi} {and} \bibinfo{person}{Helmi
  Rais}.} \bibinfo{year}{2018}\natexlab{}.
\newblock \showarticletitle{{A survey on technical threat intelligence in the
  age of sophisticated cyber attacks}}.
\newblock \bibinfo{journal}{\emph{Computers \& Security}}  \bibinfo{volume}{72}
  (\bibinfo{date}{1} \bibinfo{year}{2018}), \bibinfo{pages}{212--233}.
\newblock
\showISSN{01674048}
\urldef\tempurl%
\url{https://doi.org/10.1016/j.cose.2017.09.001}
\showDOI{\tempurl}


\bibitem[Tovar{\v{n}}{\'a}k et~al\mbox{.}(2021)]%
        {tovarnak2021graph}
\bibfield{author}{\bibinfo{person}{Daniel Tovar{\v{n}}{\'a}k},
  \bibinfo{person}{Luk{\'a}{\v{s}} Sadlek}, {and} \bibinfo{person}{Pavel
  {\v{C}}eleda}.} \bibinfo{year}{2021}\natexlab{}.
\newblock \showarticletitle{{Graph-Based CPE Matching for Identification of
  Vulnerable Asset Configurations}}. In \bibinfo{booktitle}{\emph{2021
  IFIP/IEEE International Symposium on Integrated Network Management (IM)}}.
  \bibinfo{publisher}{IEEE}, \bibinfo{address}{New York, NY, USA},
  \bibinfo{pages}{986--991}.
\newblock
\urldef\tempurl%
\url{https://ieeexplore.ieee.org/abstract/document/9463994}
\showURL{%
\tempurl}


\bibitem[Wagner et~al\mbox{.}(2019)]%
        {Wagner2019}
\bibfield{author}{\bibinfo{person}{Thomas~D. Wagner}, \bibinfo{person}{Khaled
  Mahbub}, \bibinfo{person}{Esther Palomar}, {and} \bibinfo{person}{Ali~E.
  Abdallah}.} \bibinfo{year}{2019}\natexlab{}.
\newblock \showarticletitle{{Cyber threat intelligence sharing: Survey and
  research directions}}.
\newblock \bibinfo{journal}{\emph{Computers \& Security}}  \bibinfo{volume}{87}
  (\bibinfo{date}{11} \bibinfo{year}{2019}), \bibinfo{pages}{101589}.
\newblock
\showISSN{01674048}
\urldef\tempurl%
\url{https://doi.org/10.1016/j.cose.2019.101589}
\showDOI{\tempurl}


\bibitem[Widup et~al\mbox{.}(2020)]%
        {dbir2020}
\bibfield{author}{\bibinfo{person}{Suzanne Widup}, \bibinfo{person}{Dave
  Hylender}, \bibinfo{person}{Gabriel Bassett}, \bibinfo{person}{Philippe
  Langlois}, {and} \bibinfo{person}{Alex Pinto}.}
  \bibinfo{year}{2020}\natexlab{}.
\newblock \bibinfo{booktitle}{\emph{{2020 Data Breach Investigations Report}}}.
\newblock \bibinfo{type}{{T}echnical {R}eport}. \bibinfo{institution}{Verizon}.
\newblock
\urldef\tempurl%
\url{https://doi.org/10.13140/RG.2.2.21300.48008}
\showDOI{\tempurl}


\bibitem[Xiong and Lagerström(2019)]%
        {Xiong2019}
\bibfield{author}{\bibinfo{person}{Wenjun Xiong} {and} \bibinfo{person}{Robert
  Lagerström}.} \bibinfo{year}{2019}\natexlab{}.
\newblock \showarticletitle{{Threat modeling – A systematic literature
  review}}.
\newblock \bibinfo{journal}{\emph{Computers \& Security}}  \bibinfo{volume}{84}
  (\bibinfo{date}{7} \bibinfo{year}{2019}), \bibinfo{pages}{53--69}.
\newblock
\showISSN{01674048}
\urldef\tempurl%
\url{https://doi.org/10.1016/j.cose.2019.03.010}
\showDOI{\tempurl}


\bibitem[Zhao et~al\mbox{.}(2020)]%
        {Zhao2020}
\bibfield{author}{\bibinfo{person}{Jun Zhao}, \bibinfo{person}{Qiben Yan},
  \bibinfo{person}{Jianxin Li}, \bibinfo{person}{Minglai Shao},
  \bibinfo{person}{Zuti He}, {and} \bibinfo{person}{Bo Li}.}
  \bibinfo{year}{2020}\natexlab{}.
\newblock \showarticletitle{{TIMiner: Automatically extracting and analyzing
  categorized cyber threat intelligence from social data}}.
\newblock \bibinfo{journal}{\emph{Computers \& Security}}  \bibinfo{volume}{95}
  (\bibinfo{date}{8} \bibinfo{year}{2020}), \bibinfo{pages}{101867}.
\newblock
\showISSN{01674048}
\urldef\tempurl%
\url{https://doi.org/10.1016/j.cose.2020.101867}
\showDOI{\tempurl}


\end{thebibliography}

\end{document}